# Independent control of nucleation and layer growth in nanowires

Carina B. Maliakkal*[1,2,3], Erik K. Mårtensson[2,3], Marcus Tornberg[2,3], Daniel Jacobsson[1,3,4], Axel R. Persson[1,3,4], Jonas Johansson[2,3], Reine Wallenberg[1,3,4] and Kimberly A. Dick*[1,2,3]

[1]Centre for Analysis and Synthesis, Lund University, Box 124, 22100, Lund, Sweden.

[2]Solid State Physics, Lund University, Box 118, 22100, Lund, Sweden.

[3]NanoLund, Lund University, 22100, Lund, Sweden.

[4]National Center for High Resolution Electron Microscopy, Lund University, Box 124, 22100, Lund, Sweden.

*E-mail: carina_babu.maliakkal@chem.lu.se, kimberly.dick@ftf.lth.se

**Abstract**:

Control of the crystallization process is central to developing novel materials with atomic precision to meet the demands of electronic and quantum technology applications. Semiconductor nanowires grown by the vapor-liquid-solid process are a promising material system in which the ability to form components with structure and composition not achievable in bulk is well-established. Here we use *in situ* TEM imaging of GaAs nanowire growth to understand the processes by which the growth dynamics are connected to the experimental parameters. We find that two sequential steps in the crystallization process – nucleation and layer growth – can occur on similar time scales and can be controlled independently using different growth parameters. Importantly, the layer growth process contributes significantly to the growth time for all conditions, and will play a major role in determining material properties. The results are understood through theoretical simulations correlating the growth dynamics, liquid droplet and experimental parameters.

A central challenge in crystal growth is to understand the dynamic and transient processes underlying the nucleation and growth steps. The ability to independently control these two steps would greatly expand the potential to design the structure, morphology and properties of the resulting material. Understanding the steps in crystallization is particularly important in the emerging fields of nanoscale and confined crystal growth, where a large interface-to-volume ratio enables new degrees of freedom in designing crystal properties[1,2]. Vapor-liquid-solid (VLS) growth of nanowires, which uses a nanoscale liquid melt to control nucleation and crystal growth, offers a platform to investigate heterogeneous nucleation and growth separately. For semiconductor materials, the VLS growth process enables synthesis of metastable crystal phases[3–5], metastable semiconductor alloys[6,7], extremely high dopant and impurity incorporation[8,9], and atomically-precise lattice-mismatched[10–12] and crystal-phase heterostructures[13]. Understanding the nucleation and crystallization processes at the interface in VLS-growth, and how these are influenced by the finite size of the liquid droplet, is centrally important for controlled design of these novel nanomaterials. In particular, it is critical to understand on an atomic scale the mechanisms by which the growth process is correlated to accessible experimental parameters.

The VLS process is understood to occur by dissolution of semiconductor precursor species (or their derivatives) in a liquid metal (often Au-based), followed by precipitation of the solid semiconductor after the liquid becomes supersaturated[14,15]. This process occurs in two steps: formation of a critical nucleus at the liquid-solid interface followed by layer growth across this interface[16,17]. Since the metal droplet is relatively small, the number of atoms causing supersaturation of the liquid is finite and could be consumed during the formation of a layer[18]. After the growth of a complete layer following the nucleation, there typically is an 'incubation' (waiting) period before the nucleation of the next layer[19–21]. The serial nature of the steps suggests the potential to separately access the parameters controlling them and to use this as a means to design material properties.

In this work we demonstrate using *in situ* transmission electron microscopy (TEM) imaging of GaAs nanowire growth that nucleation and layer growth can be controlled independently. By correlating the incubation time and layer growth time to the partial pressures of Ga and As precursors, we observe that the time scales for the two steps are of similar magnitude but follow entirely different trends. For most of the studied parameter space, incubation can be tuned by varying the Ga precursor partial pressure, while the layer growth can be independently controlled using the flow of the As precursor. Towards the edges of the parameter space, interactions between these species lead to more complicated trends. Simulations of the growth show that the observed effects can be understood by considering the very different steady-state compositions of the two species in the catalyst droplet during growth. Since this effect is fundamental in nature, it can be directly extrapolated to other growth conditions, growth methods and other binary semiconductor materials.

**Nanowire growth**

GaAs nanowires were grown in a Hitachi HF-3300S environmental TEM, on $SiN_x$-coated heating chips on which Au aerosol nanoparticles (~30 nm diameter) had been deposited. Trimethylgallium (TMGa) and arsine ($AsH_3$) were used as the precursors as in a typical *ex situ* metal-organic chemical vapor deposition (MOCVD) system. These precursors were supplied via capillary tubes fed through the sample holder such that the gases entered the microscope within 4 mm of the sample area. Gas flows were controlled by mass flow controllers and pressure valves, and monitored during growth with a residual gas analyzer in the exhaust gas which had been calibrated to give partial pressures at the sample. Gas flows were chosen to give $AsH_3$ partial pressure in the same range as typical *ex situ* MOCVD nanowire growth, with slightly lower TMGa flows giving average nanowire growth rates in the range of 0.1 – 1 nm/s (comparable to typical *ex situ* GaAs nanowire growth rates of about 0.5 – 5 nm/s)[22,23]. Details of the experimental setup are found in the Methods section. TMGa and $AsH_3$ were supplied simultaneously to initiate nanowire growth, after which a suitable nanowire was selected based on its orientation relative to a hole in the supporting $SiN_x$ film and to other nanowires, in order to follow the growth as parameters were varied. Nanowire growth was imaged continuously with 50 ms exposure time per image using conventional parallel beam TEM.

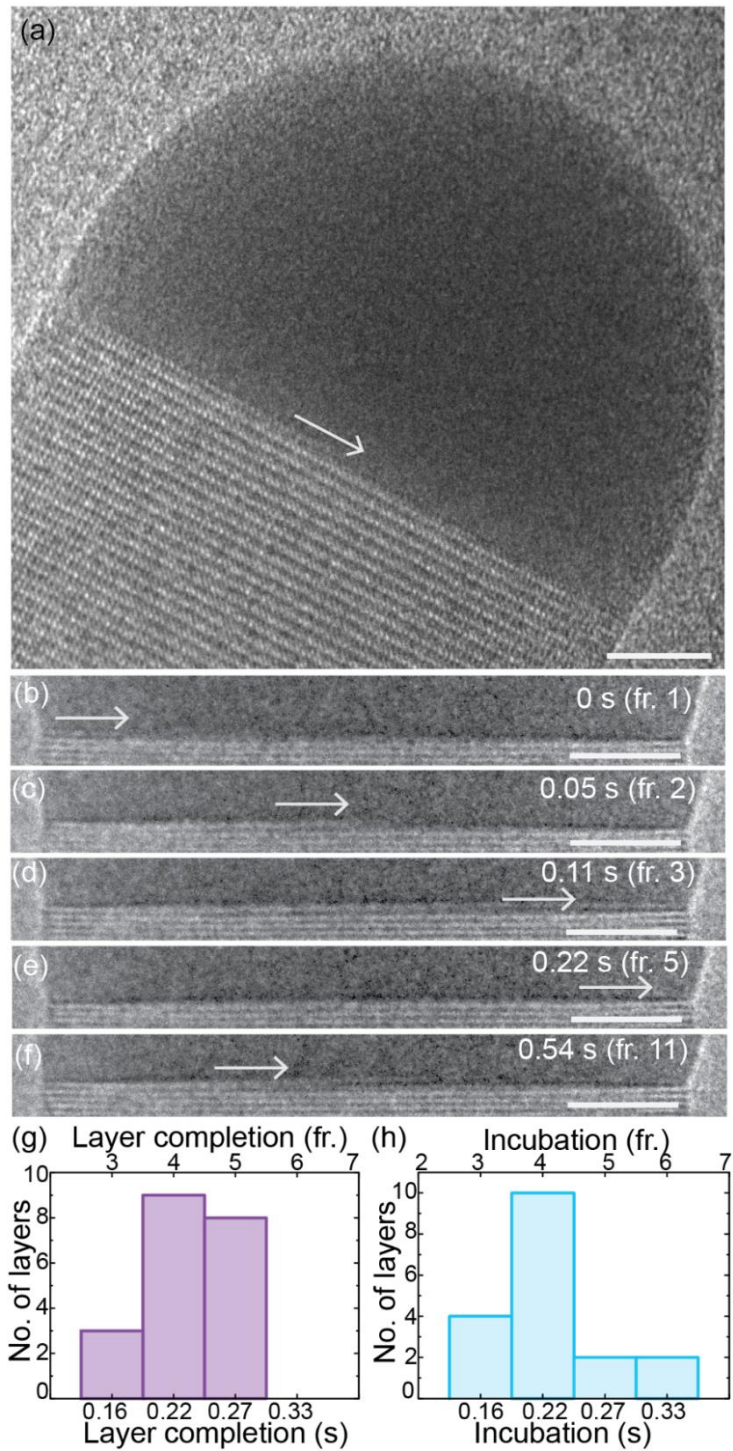

Fig. 1. **Layer growth in GaAs nanowires**: (a) TEM image of a growing GaAs nanowire NW along with the catalyst. A partially grown layer can be observed in this image (indicated by the arrow). (b)-(f) Sequence of frames (cropped) showing layer growth from a recorded video of another nanowire. The frame number (fr.) and time elapsed from the image in (b) is denoted on the top right of each section. The layer completion time, i.e. the time between the starting and ending of this layer corresponds to 4 frames (i.e. 0.22 s). After the ending of a layer in frame 5 (e) a new layer starts only in frame 11 (f). So the 'incubation time' or the waiting time in this example is 6 frames (i.e. 0.33 s). Scale bars indicate 5 nm. (See also Supplementary Video 1.) (g)-(h) Histograms of layer completion time and incubation time, respectively, of some layers grown at the same growth condition. (The frame rate for the video is 18.3 fps on an average.)

We observe that growth occurs layer-by-layer across the nanowire interface (Fig. 1 a-f), consistent with previous reports[20,21,24]. There is also a measurable 'incubation time' between successive layer growths, which we interpret as a waiting period where the droplet accumulates enough material to overcome the nucleation barrier for a new layer. This interpretation implies that there is a 'nucleation step' during which the critical nucleus (and potentially a small part of the layer) forms. This step is extremely fast and not visible in such experiments, and so will not be considered in this analysis. In the following we use the terms 'layer completion time' for the time observed for completion of each GaAs layer, and 'incubation time' as the time between completion of one layer and the starting of the next layer. Interestingly, we observe that the order of magnitude of the incubation and layer completion times is similar, with layer completion times ranging from 0.1 - 0.5 s, and incubation times ranging from about 0.2 to 5 s; an example is shown in Fig. 1 (g), (h). This is surprising, since most models on nanowire growth have assumed the layer completion time to be negligible compared to incubation time[19,25–27]. *In situ* investigations have in some cases demonstrated slow layer growth (and relatively insignificant incubation time)[20,21,28]; however, these experiments were conducted under conditions yielding very low growth rates, and hence are generally not considered to be representative of typical *ex situ* growths. Our observations suggest that, for conditions comparable to typical *ex situ* nanowire growth, both steps – incubation and layer completion – contribute in a significant way to the overall growth process.

**Kinetics as a function of As-precursor supply**

In order to understand the mechanisms controlling the incubation and layer growth, we first discuss the dependence of growth kinetics on the $AsH_3$ flow at 420 °C and TMGa pressure of $13 \times 10^{-5}$ Pa (Fig. 2). $AsH_3$ partial pressure was initially set to 1 Pa (see also Supplementary Video 2) and subsequently increased monotonically. After conducting growth at the highest $AsH_3$ pressure (5.6 Pa), the $AsH_3$ flow was reduced again to 1 Pa to verify reproducibility of the observations. The time for completing each layer (purple squares) decreases with increasing $AsH_3$ supply, indicating that layer completion is restricted by the amount of arsenic present in the catalyst. This suggests that there are not sufficient (excess) As atoms present in the droplet at any instant to form a complete layer. This is not surprising since the solubility of As in the Au-Ga alloy is very low, and experimental reports[29,30] and theoretical predictions[31] show very low concentration of As species in the catalyst. Incubation time is plotted as a function of the $AsH_3$ partial pressure as cyan circles in Fig. 2; we see that

incubation, unlike layer completion, increases slightly as we increase the As precursor pressure

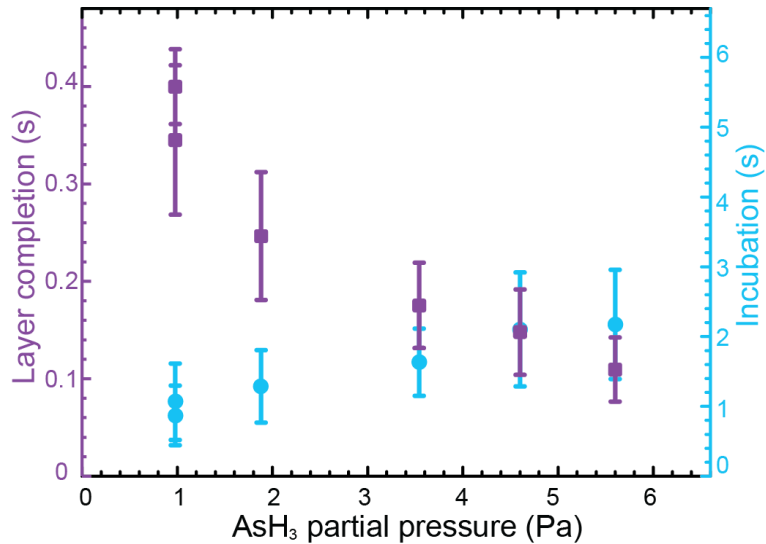

Fig. 2. **Incubation and layer completion as a function of AsH$_3$**: The average layer completion time (purple squares, axis on left side) and incubation time (cyan circles) for growing a layer is plotted as a function of the As-precursor pressure. The layer growth of each individual layer is faster at higher AsH$_3$ flow. Error bars denote standard deviation across the measured 10 or more events.

. **Kinetics as a function of Ga-precursor supply**

Next we discuss the effect of TMGa pressure while keeping the AsH$_3$ constant (1 Pa) at the same temperature of 420 °C (Fig. 3). We observe that incubation time decreases with increasing TMGa flow (i.e. nucleation occurs after a shorter average waiting time). This trend is strongest at low TMGa flow, while for high TMGa flow, the incubation time is very brief, and there is very little change with increasing flow. There is a large variability in incubation times measured for different layers growing at low TMGa flow, resulting in large error bars; for higher flows however, the spread is very small (not visible on the scale of the figure). On the other hand, the layer completion time does not depend in an obvious way on TMGa flow for most of the studied TMGa flows (9 - 60 x10$^{-5}$ Pa). For the very lowest TMGa flow investigated (9x10$^{-5}$ Pa), there is a slight increase in layer completion time, suggesting that for this extremely low Ga-supply, there might not be sufficient excess Ga atoms to form one complete layer. Finally, we observe that for a significant part of the parameter space covered in the measurement (TMGa > 25x10$^{-5}$ Pa), the layer completion time is actually longer than the incubation time. We note that this experiment was conducted on a different nanowire and day, and so the absolute values of the incubation time and layer completion time for the same 'set' experimental parameters differ slightly between experiments. This may be related to differences in local environment at the nanowire or the pumping efficiency of the instrument on different days; however, the observed qualitative trends are reproducible.

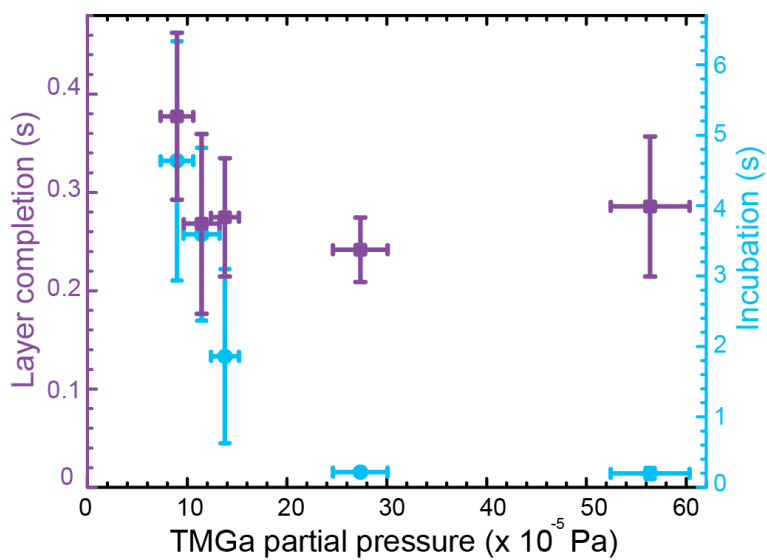

Fig. 3. **Incubation and layer completion as a function of TMGa**: Layer completion time (purple squares) and incubation time (cyan circles) are plotted as a function of the Ga-precursor flow at fixed AsH$_3$ partial pressure of 1 Pa. With increasing Ga-precursor flux, the incubation time decreases, indicating that the nucleation of each new layer is controlled by the Ga supply to the catalyst particle. For TMGa pressure above 10 x 10$^{-5}$ Pa, the layer completion time stabilizes at a non-zero value indicating that layer growth is limited by the As availability. Error bars denote standard deviation across the measured 10 or more events.

**Simulation of the growth of layers in GaAs nanowires**

The observations show that under typical growth conditions nucleation of a new layer is controlled by the Ga species, while the arrival of As species controls the layer formation. To understand this better, we conducted Monte Carlo simulations of the nanowire growth based on mass transport and nucleation theory (details are found in Methods and Supplementary

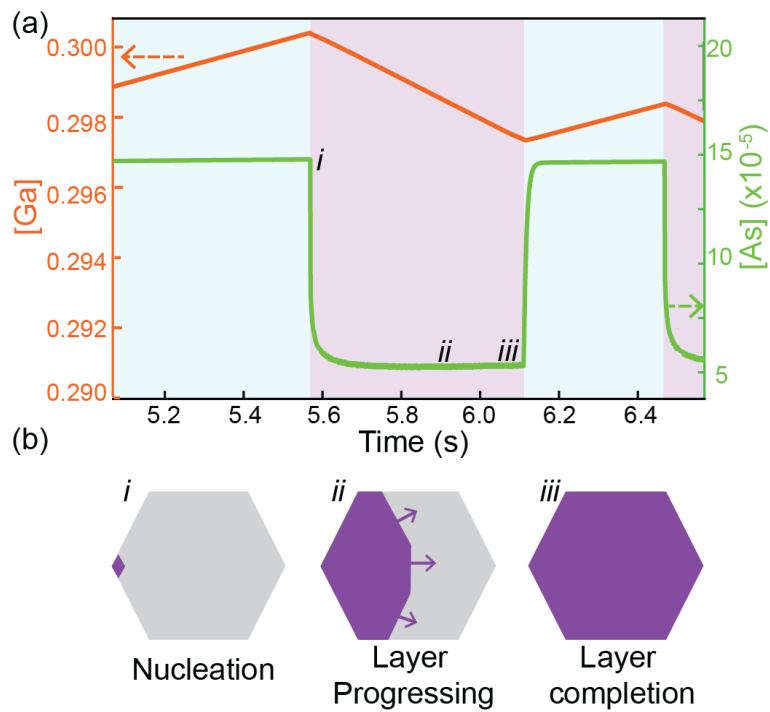

Fig. 4. **Simulation of the layer growth process in GaAs nanowires**: (a) Simulation of the Ga and As concentration in the Au-Ga-As catalyst droplet as a function of time. The collection and depletion of (excess) Ga is relatively slower than As. The As concentration is several orders of magnitude lower than Ga concentration. Different stages of layer growth (nucleation, progression of a layer and completion of a layer) are denoted in the figure as i, ii and iii respectively. (b) A schematic progression (not simulated) of these steps is shown for illustrative purposes; this schematic is not intended to display the real geometry of the growing layer, which evolves in a complex way as in Ref. [21].

Information section 1). An illustrative example of the simulated growth is shown in Fig. 4. The Ga and As concentrations in the catalyst droplet (and several other related parameters) change in a cyclic way: increase until a critical supersaturation is reached, then decrease when nucleation occurs (*i*) and a layer forms (*ii*). Following completion of a layer (*iii*), the concentrations again increase. Arsenic has a high vapor pressure and very low equilibrium solubility in Au[31]; droplet supersaturation is thus very sensitive to the addition of As atoms, and As concentration rapidly equilibrates with the ambient $AsH_3$ pressure. Gallium on the other hand easily forms metallic liquid alloys with Au[32], and the Ga concentration in the droplet is high during nanowire growth (measured experimentally to be on the order of 25-55 atomic %)[29], with additional Ga atoms having a relatively much smaller effect on the droplet supersaturation. In order for nucleation to occur, the species in the droplet must exceed the nucleation barrier as determined by the supersaturation of Ga and As in the droplet relative to the GaAs crystal. When the As concentration has equilibrated with the vapor, the probability of overcoming this barrier is typically low enough to prevent nucleation. While the As concentration remains flat, the Ga concentration continues to increase (Fig.4 (a)) resulting in a higher nucleation probability. Thus, the nucleation of a new layer is controlled by Ga for most of the parameter space (further details on the parameters controlling the different steps for different regimes are found in Supplementary Information section 2). When a nucleation event occurs the number of available As atoms drops rapidly and is insufficient to form a complete monolayer, and so layer growth will primarily be controlled by the rate at which As atoms arrive.

Monte Carlo simulations were performed for about 35 cycles of growth for each precursor flow to find the average layer completion and incubation times (Fig. 5). The error bars represent the standard deviation among 'grown' layers in the simulation. We can see that the simulation reproduces the trends observed experimentally (Fig. 2, 3): Layer completion time decreases strongly with increasing As partial pressure while showing a very weak, almost flat, dependence on Ga partial pressure. Incubation on the other hand decreases strongly with increasing Ga partial pressure, while increasing slightly with increasing As partial pressure. The observation of a small increase in incubation with As pressure in the simulation allows us to understand its origin: some of the arriving Ga species accumulate in the catalyst during the layer completion, so the corresponding incubation time (after the layer growth is completed) is slightly longer when the layer completion time is shorter. Finally, the simulation also shows that the spread in incubation times between individual layers is much larger for low Ga partial pressure, and that layer-to-layer variability in completion time is very small for all conditions.

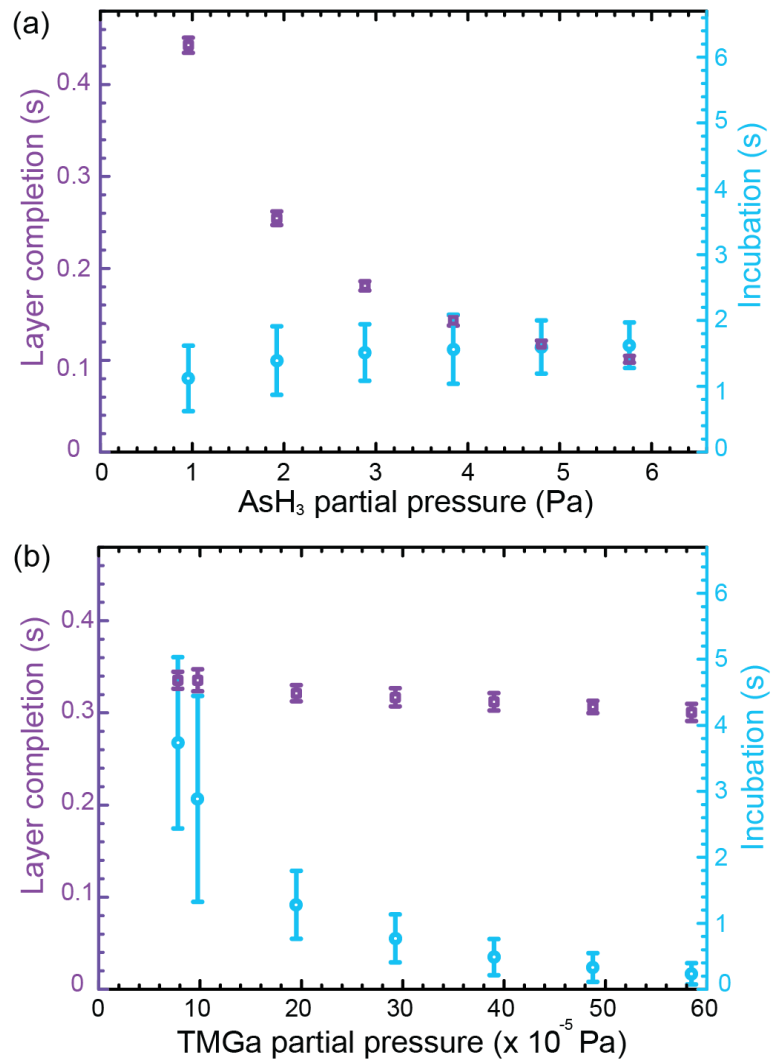

Fig. 5. **Simulation of incubation and layer growth as a function of precursor pressures**: Layer completion time and incubation time obtained from Monte Carlo simulations for (a) the $AsH_3$ series and (b) TMGa series. The error bars represent the standard deviation among 'grown' layers.

**Discussion**

The results here show that the time scales for incubation and layer completion are comparable, and hence nanowire growth cannot be completely understood by considering only one species or only one process. Overall growth rate depends more strongly on the Ga species than As species for most of the accessible parameter space (in our experiments and in comparable *ex situ* growths). However, the observation that incubation/nucleation and layer growth are primarily controlled by different experimental parameters (Ga and As precursor partial pressures, respectively) means that under appropriate conditions it is possible to independently control the two steps. The evolution of the droplet composition during the two steps will control such properties as composition, uniformity and crystal structure; clearly understanding the time scales for the two events thus provides a greater freedom to design the material properties. Crystal structure for instance is generally controlled by nucleation, assuming the absence of structural defects during layer growth or post-growth structural transformations. On the other hand, compositional uniformity and impurity/dopant incorporation can also strongly depend on the layer growth step. For example, incorporation of impurities by step trapping, which can result in very high concentrations of dopants or other trace elements, is directly proportional to the layer-propagation rate rather than overall growth rate[6,33]. Similarly, the performance of heterostructure devices such as tunnel field-effect transistors depends critically on the composition of the heterojunction[34], which will be sensitive to the relative rates of both the incubation and layer growth processes.

As noted earlier, parameters in this study were chosen to give nanowire growth rates comparable to the lower end of typical growth rates observed for *ex situ* MOCVD, achieved by using similar $AsH_3$ pressures and lower TMGa pressures. This indicates that *ex situ* incubation times will tend to be slightly shorter than observed here, while layer completion times will be similar – together suggesting that the similar time scales and distinct trends observed for these events will also be applicable for most *ex situ* MOCVD nanowire growth. Since the effects are not related to the types of precursors used, they should apply equally well to other growth techniques (such as molecular beam epitaxy). Moreover, although this study focuses on gold-seeded GaAs nanowire growth, the key insights are not restricted to this system. The fact that Ga readily alloys with the metallic catalyst while As incorporation into the catalyst is minimal is the underlying reason why Ga and As control growth in very distinct ways. Most other III-V and II-VI binary semiconductors are also composed of one metallic (group II or III) element and one nonmetallic (group V or VI) element, which in most cases exhibit similar alloying trends with gold or other typical transition metal catalyst[32,35]. It is thus reasonable to hypothesize that it will be possible to independently control layer nucleation (typically with the group II/III precursor) and layer growth (typically with the group V/VI precursor) for many of these material combinations.

**Summary and Conclusions**

Understanding the mechanistic processes controlling nucleation and crystal growth is key to designing materials with desired properties. In particular, interface-dominated, nanoscale systems offer new opportunities to control the crystallization process and resulting properties. In this study we have investigated *in situ* the nucleation and layer growth processes in VLS growth of GaAs nanowires. We observe that the nucleation (controlled by an incubation process) and layer growth processes can be independent, occur on similar time scales and are controlled by different parameters for most of the growth parameter space. This is a consequence of the confined multi-phase nature of the VLS process: Although material is supplied via the gas phase, the kinetics of the process are controlled by the catalyst droplet, which is very small. This means that the nucleation and initial layer growth can significantly change the composition of the droplet, depleting it of supersaturating atoms. For binary materials composed of two species that interact very differently with the catalyst droplet material, this opens up a broad parameter space in which nucleation and layer growth are independently controlled. Specifically, nucleation can be controlled by the species that easily alloys with the catalyst (such that each atom changes the chemical potential very little) while layer growth is controlled by the other species, for which very few atoms are required to supersaturate the droplet. The similar time scales of these two steps indicate that both must be considered in order to understand the nanowire growth process and to control the material properties. Controlling the parameters dominating the independent steps offers an opportunity to engineer devices with controlled properties, such as crystal structure, composition and impurity incorporation.

**Methods**

**Data acquisition and presentation:** Videos of atomic resolved images were recorded with an AMT XR401 sCMOS camera with an exposure time of 50 ms per image. Continuously recording results in videos with about 18.3 frames per second (fps) on average. The TEM images shown here were extracted from these videos and processed. The incubation time and layer completion time are measured from the videos. The error bars in the plots of incubation time and layer completion time represent the standard deviation among multiple layer growth events.

*In situ* **growth:** GaAs nanowires were grown in a Hitachi HF-3300S environmental transmission electron microscope (ETEM) with a cold field emission gun and a CEOS B-COR-aberration-corrector. The wires grew along the <111> direction (cubic notation, or equivalent <0001> direction in hexagonal notation). Blaze software supplied by Hitachi was used to control the local sample temperature using Joule heating in a constant resistance mode. The heating chips used had thin $SiN_x$ windows on which nanowire growth could be followed; some contained holes so that growth could be viewed without $SiN_x$ in the background. The ETEM was connected to a gas handling system with the MOCVD gases. Further details of the experimental setup can be found in Ref. [29].

**Precursor supply:** $AsH_3$ and TMGa were used as the precursors. The $AsH_3$ flow was controlled exclusively by the mass flow controller. The TMGa was supplied via a bubbler maintained at -10 °C with $H_2$ bubbled through it. A portion of the TMGa/$H_2$ flow passing the mass flow controller was bypassed to the vent line to restrict the TMGa pressure reaching the microscope. The precursor fluxes sent to the ETEM were monitored with a residual gas analyzer (SRS RGA 300) using mass spectrometry. The precursor flows were calibrated to find the partial pressure at the sample. More details on the calibration methodology is given in Ref. [29].

**AsH$_3$ series experiments:** The AsH$_3$ flow was varied at constant TMGa pressure (13.1x10$^{-5}$ Pa) and temperature (420 °C). The AsH$_3$ supply partial pressure was initially set to 1 Pa, then increased in steps to 5.6 Pa and finally decreased back to 1 Pa.

**Ga series experiments:** The TMGa partial pressure was varied at a fixed AsH$_3$ supply (~1 Pa) and temperature (420 °C). This experiment was started at a TMGa pressure of 11x10$^{-5}$ Pa, and successively set to 9, 13, 27 and finally 55 x 10$^{-5}$ Pa.

**Specifications about the model:** The Monte Carlo simulation model is based on the work in Ref. [31], which uses the effective impingement rates of As and Ga as the main input parameters. A complete description of the model is found in Supplementary Information section 1; here we summarize the simulation process and how it was used in this study. Time steps of 1 ms are taken and in each step, the composition of the seed particle is updated based on the effective impingements, the evaporation of As and Ga and potentially on Ga-As pairs incorporated in the layer growth. The nucleation rates, which depend on the properties of the seed (such as supersaturation), are calculated using classical nucleation theory, modified to take droplet depletion[18] into consideration. Whether or not a nucleation event occurred in the time step is decided by random numbers and the nucleation rates. Once a nucleation event has occurred (and layer growth is not yet completed), Ga-As pairs are moved from the liquid seed to the solid layer in each time step, as long as the supersaturation is high enough for it to be energetically favorable to grow the nucleus further. The layer completion time and incubation time are thus obtained for each cycle. The values shown in Fig. 5 are obtained by averaging over multiple such simulation cycles (about 35 cycles for each set of growth conditions). The error bars in the modelled value denote the standard deviation among the multiple 'layer growth' events in the simulation. In this model we assume that the Ga (or As) flux into the catalyst is directly proportional to the TMGa (or

AsH$_3$) partial pressure. Due to the slight difference between the absolute values measured for the two different experiments, the proportionality constant between AsH$_3$ partial pressure and the As flux into the particle was adjusted. (As a consequence, the 1 Pa AsH$_3$ pressure used for the Ga series experiment was set in the simulation to be equivalent to 1.5 Pa AsH$_3$ of the AsH$_3$ series.) The proportionality constants will be affected by for instance V/III or pressure dependent precursor decomposition, TMGa-dependent Ga surface diffusion, effect of local environment near the nanowire, and the spatial variation of the gas flow (in terms of fluid dynamics); therefore, these effects are not explicitly accounted for.

**Acknowledgments**

We wish to acknowledge support from the Knut and Alice Wallenberg Foundation, NanoLund and the Swedish Research Council.

**Author contributions**

CBM, DJ, MT and ARP performed the experiments. Data analysis was done by CBM. EKM performed the theoretical simulations with input from JJ. KAD and RW coordinated the project. All authors discussed the results and contributed to the concepts discussed in this article.

**Competing financial interests**

The authors declare no competing financial interests

# Supplementary Information: Independent control of nucleation and layer growth in nanowires

## 1. Specifications about the simulations

The model used for the simulation is based on the work of Mårtensson *et al.*[1], modified to treat nucleation and layer-growth as two separate events. The model follows the mass transport of material through the seed particle over time; to achieve this, time is discretized into steps of dt=1 ms. Within each time step, the transfer of material between the vapor and liquid phases is treated, as well as any transfer of material from the liquid to the solid phase via either nucleation or layer-growth.

The net vapor-liquid flow ($J_{net}$) is calculated using:

$$J_{net} = \frac{1}{\sqrt{2\pi m k_B T}}(P_{in} - \eta N P_{out})$$

Here, m is the atomic mass of the species, $k_B$ is Boltzmann's constant, T is the growth temperature, $P_{in}$ is the pressure of the impinging growth species (chosen at the start of a simulation), $P_{out}$ is the vapor pressure of the species in the seed particle, $\eta$ is a factor describing the efficiency of the evaporation set to $2\times10^{-4}$ after fitting, and N is the number of atoms per evaporation event (here 4 for $As_4$ and 1 for Ga). The vapor pressure is calculated as:

$$P_{out} = P_b \, exp((\Delta H_{vap})/R(1/T - 1/T_b)) \exp((\mu - \mu_0)/kT)$$

The pressure at which the element boils, $P_b$, is first scaled from the boiling temperature ($T_b$) to the current temperature (420 °C) using the enthalpy of vaporization ($\Delta H_{vap}$), and the Clausius-Clapeyron relation. The vapor pressure is then adjusted to take the environment into consideration using the difference between the chemical potential of a pure species ($\mu_0$) and that of the species in the seed particle ($\mu$). Values for $P_b$, $T_b$ and $\Delta H_{vap}$ for Ga are easily found; however, since As sublimates under normal circumstances the values found in literature vary. Here, we used the lowest enthalpy of forming $As_4$ in Ref.[2] as an estimate. The boiling point was set to 36.2 bar at 1090 K for As. The net flow, $J_{net}$, gives the net number of atoms per time and area and to get the net number of atoms which are incorporated into the seed, $J_{net}$ is multiplied by dt and the area of the seed as in Ref.[1].

The chemical potentials needed are calculated from the molar Gibbs free energy (g) of a ternary alloy described by:

$$g = \sum_{i=1}^{3} X_i \mu_{0,i} + \sum_{i=1}^{3} X_i RT ln(X_i) + \sum_{i=1}^{2} \sum_{j=i+1}^{3} X_i X_j \sum_{v=0}^{2} L_{i,j,v} (X_i - X_j)^v$$

The first term sums the unary chemical potentials based on the molar fraction of each species ($X_i$), the second term is the entropy of mixing (with R being the gas constant), and the third term takes binary interactions into account in the form of the Redlich-Kister expansion. For the As-Au-Ga system, no ternary interaction parameter was found in literature, and for this reason it was not considered here. The temperature-dependent interaction parameters $L_{i,j,v}$ are taken from literature (As-Au[3,4], As-Ga[5], Au-Ga[6]), and so are the expressions for the unary chemical potentials[7]. The chemical potential for each species in the mixture is then calculated from the derivatives of G as:

$$\mu_i = \frac{\partial G}{\partial N_i}, \qquad G = g \sum_j N_j / N_{Av}, \qquad X_i = \frac{N_i}{\sum_j N_j}$$

Here, $N_i$ represents the number of atoms of each component in the seed particle and $N_{Av}$ is Avogadro's constant. This framework allows us to calculate the chemical potential for each species, which is needed for the vapor-liquid transfer.

Next, we can evaluate the energy barrier of forming a stable nucleus. From previous experiments[8], we assume that the experimentally grown nanowires were predominantly wurtzite (WZ), and for this reason we do not include the zinc blende (ZB) phase in the model. Because of the small radius of the nanowire (measured to be 16 nm), we chose to include the energy of the liquid phase into the nucleation model. The motivation to take the energy of the liquid seed into consideration is the depletion effect, as presented by in Dubrovskii.[9] This effect refers to the influence that removing III-V pairs to form or grow a nucleus has on the liquid seed particle, which becomes more significant as the seed particle becomes smaller. To describe the liquid to solid transition, we consider the Gibbs free energy ($G_{sys}$) of the liquid seed particle and a nucleus or a layer consisting of i WZ pairs, which we describe as:

$$G_{sys}(N_{As}, N_{Ga}, i) = N_{As}\mu_{As} + N_{Au}\mu_{Au} + N_{Ga}\mu_{Ga} + i(\mu_{GaAs} + \psi) + \sqrt{i}\phi$$

The first three terms give the energy of the liquid phase, with $N_{Au}$ set to an estimated value of 600000 atoms and $N_{As}$, $N_{Ga}$ dynamically changing over time. The last two terms give the energy of the nucleus or incomplete layer of the WZ phase. The chemical potential of the solid WZ phase is calculated by first calculating the chemical potential of ZB, $\mu_{GaAs}$, using the data in Ref.[10], and the difference in bulk cohesive energy between ZB and WZ, $\psi$, which is taken as 23.1 meV/pair[11]. The chemical potentials are in units of J/atom for the species and J/pair for the binary. Here, $\phi$ is the average surface energy of the nucleus and is calculated as $\phi = C\Gamma\sqrt{\Omega_s h}$. The constant C relates to the shape of the nucleus, and is given by the ratio of the perimeter to the square root of the surface area of the top of the nucleus. For a triangular shape assumed here, C is approximately 1.86. The factors $\Omega_s$ (volume of a GaAs pair in the solid) and $h$ (height of a bilayer) are assumed to be the same for both crystal phases and are calculated from bulk ZB values. The effective surface energy of the perimeter, Γ, is based on the model by Glas *et al.* of triple phase line nucleation,[12] and is calculated as $\Gamma = (1-x)*\gamma_{SL} + x(\gamma_{SV} - \gamma_{LV}\sin\beta)$, using the interface energies for the solid-liquid $\gamma_{SL}$ fitted to be 0.1 J/m², the solid-vapor taken from Tornberg *et al.*[13] and the liquid-vapor times the sine of the contact angle as described in Ref.[1]. The variable x refers to the fraction of the nucleus which is in contact with the vapor, and for a triangular nucleus this equates to 1/3. However, since the expression for the effective surface energy is used to describe both the initial nucleus and the subsequent growth of the bilayer (which in the end form a complete layer where the entire perimeter is in contact with the vapor), the value of x was chosen to vary linearly with the coverage of the nucleus from 1/3 at the initial nucleation to 1 at a complete layer.

With the energy of the system described above, the change in Gibbs free energy of adding a new GaAs pair is given by:

$$\Delta G^{+1}(N_{As}, N_{Ga}, i) = G_{sys}(N_{As}-1, N_{Ga}-1, i+1) - G_{sys}(N_{As}, N_{Ga}, i)$$

We note here that many variables used to calculate $G_{sys}$ depend on the number of Ga and As atoms in the seed particle (such as the chemical potentials), and these variables are re-calculated after removing a Ga-As pair from the liquid.

In the incubation periods, we are interested in the rate at which a stable nucleus is formed, which depends on the size at which it is favorable for a nucleus to grow. The critical size of the nucleus ($i^*$) is defined as the smallest $i$ where $\Delta G^{+1}(N_{As}-i, N_{Ga}-i, i) < 0$. This gives the effective nucleation barrier as:

$$\Delta G^* = \sum_{i=0}^{i^*-1} \Delta G^{+1}(N_{As}-i, N_{Ga}-i, i)$$

From this, the nucleation rate is calculated as per the classical nucleation theory as:

$$\omega = \omega_0 \exp(-\Delta G^*/kT)$$

where $\omega_0$ is the pre-exponential factor which was set to $10^9$ s$^{-1}$. The value for the pre-exponential factor was fitted to give a variance of the incubation times in the As series of a similar scale as measured in the experiments. For determining whether or not a successful nucleation occurred in the current time step, the rates were converted to probabilities using the cumulative exponential distribution function, and the outcome was chosen using random numbers as described in Ref.[1]. When a successful nucleation occurs, the time is noted, and $i^*$ Ga-As pairs are removed from the liquid and added to a now stable, growing new bilayer.

The simulation changes to layer-growth mode once a stable nucleus has formed. At this stage, we assume that the liquid is always in equilibrium with the nucleus, meaning that the attachment rate of GaAs pairs to the nucleus is not limited by kinetic considerations in the liquid. In each time interval, the composition of the seed is first updated based on the impingement as before. Then, to find how much the layer grows, the inequality $\Delta G^{+1}(N_{As}, N_{Ga}, i) < 0$ is considered. If this inequality holds, it is energetically favorable for the layer to grow, and one Ga-As pair is removed from the liquid ($N_{As/Ga} \rightarrow N_{As/Ga} - 1$) and added to the nucleus ($i \rightarrow i + 1$), which lowers the supersaturation of the liquid seed particle. This layer growth is repeated until the inequality no longer holds, at which point the system moves on to the next time interval. This iterative process is carried out until a complete layer has formed. Once the layer is completed, the time is noted and the simulation resets ($i \rightarrow 0$) and goes back to incubation mode and tries to grow the next layer. The times at which nucleations occurred and layers were completed were then used to calculate the mean and standard deviation of the incubation and layer completion times presented in the main text.

## 2. Different regimes in III-V nanowire growth

Experimental reports typically refer to 'V-limited' and 'III-limited' growth, assuming that if one species is in excess, the growth is controlled by the other species. The observation that incubation/nucleation and layer growth can be controlled by different species means however that the overall picture is more complex. On the basis of our understanding we identify the following regimes (summarized in Table 1):

*(i)* As-limited regime: At low $AsH_3$ flow and low V/III ratio, As limits both incubation (which will be very short, approaching zero) and layer-growth (which can become very long) and hence growth becomes V-limited. In this regime Ga will rapidly accumulate in the droplet, and so the parameter space for stable growth is very narrow[14]. (Moderately high Ga accumulation in the seed particle can increase the droplet contact angle resulting in truncation corners that could in turn modify the layer growth dynamics[15]. The experiment discussed here are performed intentionally at conditions where a truncation corner is not present. Hence, the effects of truncation on the layer growth kinetics are not considered in these simulations.)

| Regime | As-limited | bi-limited | Semi-Ga-limited | Quasi-Ga-limited | True Ga-limited |
|---|---|---|---|---|---|
| Conventional terminology | 'As-limited' | | 'Ga limited' | | |
| Layer completion limited by | As | As | As | As, Ga | - |
| Nucleation/Incubation limited by | As | As & Ga | Ga | Ga | Ga |
| Overall nanowire growth rate determined by | As | As & Ga | mainly Ga | mainly Ga | Ga |

Table 1. Growth regimes: Different growth regimes can be identified according to what species determines the layer-growth and nucleation processes. The different regimes are ordered here in a way that on the left column As flow is low (i.e. Ga flow is relatively high) and on the right column it reverses to high As flow (i.e. Ga flow is relatively low).

*(ii)* Bi-limited regime: At slightly higher $AsH_3$ flow than the As-limited regime, there would exist an intermediate regime where the steady-state Ga and As concentrations in the droplet are correlated and incubation time determined by both flows, while layer-growth is still controlled by $AsH_3$.

*(iii)* Semi-Ga-limited regime: With further increase in $AsH_3$ flow the V/III ratio is high enough that the As concentration in the droplet rapidly equilibrates with the vapor; consequently TMGa flow will control nucleation; but $AsH_3$ will still control layer-growth. Almost the entire parameter space accessible *in situ* falls in this regime.

*(iv)* Quasi-Ga-limited regime: At even higher V/III ratio (with moderate or low Ga supply, but high $AsH_3$ flow) there exists an intermediate regime where both Ga and As availability at the growth-front determines the layer completion time. The slight increase of layer-completion time observed at the lowest TMGa flux studied experimentally might be an indication of this regime.

*(v)* True-Ga-limited regime: At low TMGa flow and very high absolute $AsH_3$ flow the incubation time would be limited by Ga. The layer completion could either be instantaneous or be controlled by Ga (assuming diffusion of As through the catalyst is fast enough).

For conditions where both the incident Ga and As precursor flux are very high the simulation predicts layer completion time tending to zero (due to the surplus supply of atoms stored in the liquid), but incubation time remains non-zero. However, in reality growth rates could be determined by diffusion of As or Ga in the liquid, or kinetic factors such as the actual crystallization time, which are not considered in the simulation.

Arsenic pressures of over 40 Pa would be a minimum requirement according to rough estimates from these simulations to have more As atoms in the seed particle than in one complete bilayer and thus form a complete layer instantly (at 420 ºC). $AsH_3$ partial pressures close to 40 Pa is possible in *ex situ* MOCVD but would be rarely used for GaAs nanowire growth[14]. Most earlier reports on *ex situ* growths where reported for 0.5 – 6 Pa range, so we anticipate that layer growth would not be instantaneous under most common MOCVD conditions.

In the experiments reported in this paper the growth was never observed to be either exclusively group V-limited or group III-limited. V-limited growth is achievable experimentally, but since Ga accumulates in the droplet with time, nanowire stability is limited and the parameter range for successful growth is very small. True III-limited growth would necessarily require much higher absolute $AsH_3$ pressure, so that the concentration of As atoms in the droplet at nucleation would be sufficient to form one complete layer of GaAs. A true III-limited growth might be achievable for certain conditions in *ex situ* MOCVD.